\documentclass{emulateapj}

\slugcomment{Astrophysical Journal Letters, submitted 2004 March 19, accepted 2004 May 5, scheduled for GALEX special issue}

\shorttitle{RR Lyrae stars in the UV}
\shortauthors{WHEATLEY et al.}

\begin{document}


\title{Large-Amplitude Ultraviolet Variations in the RR Lyrae Star ROTSE-I~J143753.84+345924.8}


\author{
Jonathan M. Wheatley,\altaffilmark{1}
Barry Y. Welsh,\altaffilmark{1}
Oswald H. W. Siegmund,\altaffilmark{1}
Yong-Ik Byun,\altaffilmark{2}
Sukyoung Yi,\altaffilmark{3}
Young-Wook Lee,\altaffilmark{2}
Barry F. Madore,\altaffilmark{4,5}
Maurice Viton,\altaffilmark{6}
R. Michael Rich,\altaffilmark{7}
Luciana Bianchi,\altaffilmark{8}
Tom A. Barlow,\altaffilmark{9}
Jose Donas,\altaffilmark{6}
Karl Forster,\altaffilmark{9}
Peter G. Friedman,\altaffilmark{9}
Timothy M. Heckman,\altaffilmark{8}
Patrick N. Jelinsky,\altaffilmark{1}
Roger F. Malina,\altaffilmark{6}
D. Christopher Martin,\altaffilmark{9}
Bruno Milliard,\altaffilmark{6}
Patrick Morrissey,\altaffilmark{9}
Susan G. Neff,\altaffilmark{10}
David Schiminovich,\altaffilmark{9}
Todd Small,\altaffilmark{9}
Alex S. Szalay,\altaffilmark{8} and
Ted K. Wyder\altaffilmark{9}}

\altaffiltext{1}{Experimental Astrophysics Group, Space Sciences Laboratory, University of California, Berkeley, CA 94720;
 wheat@ssl.berkeley.edu, bwelsh@ssl.berkeley.edu, ossy@ssl.berkeley.edu, patj@ssl.berkeley.edu.}

\altaffiltext{2}{Center for Space Astrophysics, Yonsei University, Seoul
120-749, Korea;
 byun@obs.yonsei.ac.kr, ywlee@csa.yonsei.ac.kr.}

\altaffiltext{3}{Department of Physics, Oxford University, Keble Road, Oxford OX1 3RH, UK;
 yi@astro.ox.ac.uk.} 

\altaffiltext{4}{Observatories of the Carnegie Institution of Washington,
813 Santa Barbara St., Pasadena, CA 91101}

\altaffiltext{5}{NASA/IPAC Extragalactic Database, California Institute
of Technology, Mail Code 100-22, 770 S. Wilson Ave., Pasadena, CA 91125;
 barry@ipac.caltech.edu.}

\altaffiltext{6}{Laboratoire d'Astrophysique de Marseille, BP 8, Traverse
du Siphon, 13376 Marseille Cedex 12, France;
 maurice.viton@free.fr, jose.donas@oamp.fr,  roger.malina@oamp.fr, bruno.milliard@oamp.fr.}

\altaffiltext{7}{Department of Physics and Astronomy, University of California, Los Angeles, CA 90095;
 rmr@astro.ucla.edu.}

\altaffiltext{8}{Department of Physics and Astronomy, The Johns Hopkins
University, Homewood Campus, Baltimore, MD 21218;
 bianchi@pha.jhu.edu, heckman@adcam.pha.jhu.edu, szalay@pha.jhu.edu.}

\altaffiltext{9}{California Institute of Technology, MC 405-47, 1200 East
California Boulevard, Pasadena, CA 91125;
 tab@srl.caltech.edu, krl@srl.caltech.edu, friedman@srl.caltech.edu, cmartin@srl.caltech.edu, patrick@srl.caltech.edu, ds@srl.caltech.edu, tas@srl.caltech.edu, wyder@srl.caltech.edu.} 

\altaffiltext{10}{Laboratory for Astronomy and Solar Physics, NASA Goddard
Space Flight Center, Greenbelt, MD 20771;
 neff@stars.gsfc.nasa.gov.}



\begin{abstract}
The NASA Galaxy Evolution Explorer ($\it GALEX$) satellite
has obtained simultaneous near and far ultraviolet light curves of
the ROTSE-I Catalog
RR~Lyrae type ab variable star J143753.84+345924.8. A series of 38
$\it GALEX$ Deep Imaging Survey observations well distributed in
phase within the star's 0.56432~d period shows an AB~=~4.9~mag
variation in the far UV (1350~-~1750~\AA) band and an AB~=~1.8~mag
variation in the near UV (1750~-~2750~\AA) band, compared with only a
0.8~mag variation in the broad, unfiltered ROTSE-I ($\approx$4500~-~10000~\AA) band. 
These $\it GALEX$ UV observations are the first to reveal a large RR Lyrae 
amplitude variation at wavelengths below 1800~\AA. 
We compare the $\it GALEX$ and ROTSE-I observations 
to predictions made by recent Kurucz stellar atmosphere models.
We use published physical parameters
for the comparable period (0.57433~d), well-observed RR Lyrae star WY Antliae
to compute predicted FUV, NUV, and ROTSE-I light curves for J143753.84+345924.8.
The observed light curves 
agree with the Kurucz predictions for [Fe/H] = -1.25
to within AB~=~0.2 mag in the {\it GALEX} NUV and ROTSE-I bands, and within 0.5~mag in the FUV.
At all metallicities between 
solar and one hundredth solar, the Kurucz models predict 6~-~8~mag
of variation at wavelengths between 1000~-~1700~\AA.
Other variable stars with similar temperature variations, such
as Cepheids, should also have large-amplitude FUV
light curves, observable during the ongoing $\it GALEX$ imaging surveys.

\end{abstract}


\keywords{stars: atmospheres --- stars: individual (ROTSE-I J143753.84+345924.8) --- stars: variables: other (RR Lyrae) --- ultraviolet: stars }


\section{Introduction}
RR Lyrae stars vary in brightness primarily due
to radial pulsations, such that
contraction and expansion of their stellar surface 
produces a temperature variation that
is observed as a change in their apparent stellar magnitude.
This cyclic behavior
has been well-observed for many thousands of these
Population II stars, with an
apparent magnitude variation of $\sim$~0.3~mag when observed at
near infrared wavelengths and a variation of $\sim$~1.0~mag
at visible wavelengths \citep{ski93}.
Since the physical mechanism causing the
brightness variation is the same for each RR Lyrae star, it
follows that their absolute magnitudes are also similar and thus
they provide a useful tool for determining stellar distances. 
In addition, observation of their periodic variability can also
provide important empirical tests of stellar pulsation and
stellar atmosphere theory \citep{kur02,mih03}.
 
Although the variation in apparent magnitude in the
ultraviolet (UV) regime is far more pronounced (i.e.$>$~2 
mag), observations of RR Lyrae stars for wavelengths
$<$~3000~\AA\ are surprisingly sparse.  The first observation
of the UV light curve of an RR Lyrae star was reported by \citep{hut77}
using photometry at 1550~\AA\
gained with the $\it OAO-2$ satellite. 
Since then, similar observations
in the UV have been reported for RR Lyrae,
X Arietis, W Virginis and several other bright variables using the
$\it ANS$ and $\it IUE$ satellites \citep{bon82,bon85,boh84}.
In general, most of
these data have been recorded at $\lambda$ 
$>$ 1750~\AA\ with an incomplete sampling over the whole phase of
the stellar light curves. In this $\it Letter$ we present
observations well distributed in phase of the UV light curve 
of an RR Lyrae star, catalogued by the Robotic Optical Transient Search Experiment
(ROTSE-I) as J143753.84+345924.8 \citep{ake00}.
The Galaxy Evolution Explorer ($\it GALEX$) \citep{mar04} obtained simultaneous light curves in
the wavelength ranges (1350 - 1750~\AA) and (1750 -2750~\AA).

\section{Observations and Data Reduction}

$\it GALEX$ observations of the star ROTSE-I J143753.84+345924.8 (hereafter named R-J14+34.5),
were obtained serendipitously during a 15~d interval in 2003 July.
The 38 exposures, most between 1000 - 1700~s
duration, were part of the mission's Deep Imaging Survey program to map a
region named  NGP\_DWS\_00 near the north galactic pole. Data were collected in
the form of time-tagged
photon events in the far (1350~-~1750~\AA) and near (1750 -
2750~\AA) ultraviolet channels. These
photons were then processed by the
$\it GALEX$ science data analysis pipeline \citep{mor04}
to produce a 3840~px square image (1.5~arc sec\ px$^{-1}$) for each
of the UV detectors.
Automated Source Extractor software \citep{ber96} subsequently detected
sources and computed aperture photometry on
them, resulting in an AB magnitude \citep{oke74} in both the $\it GALEX$
far (FUV) and near (NUV) ultraviolet wavebands.  Table 1 details
the date, exposure time and $\it GALEX$ magnitude for each observation.

The ROTSE-I visible light curve consists of 76 observations recorded
over a 95~d interval, in 1999 March-June \citep{ake00}.
ROTSE magnitudes are derived from unfiltered CCD images with a broad 4500~-~10000~\AA\ bandpass
\citep{woz04}, and we note that
Blazhko amplitude modulation is not evident in the ROTSE-I data.

To estimate the distance to R-J14+34.5, we first convert the mean m$_{ROTSE}$
to an m$_V$, using the conversion factors of \citet{amr01}.
The distance to R-J14+34.5 is estimated to be 3.6$\pm$0.5~kpc, derived using 
m$_{ROTSE}$~=~13.78, m$_V$=13.64, an
absolute magnitude of M$_{V}$~=~0.6 and E(B-V)~=~0.09, taken from 
the reddening maps of \citet{sch98}.

%%\clearpage

%%\begin{figure}
%%\epsscale{0.8}
%%\plotone{f1.eps}
%%\caption{ROTSE-I and {\it GALEX} light curves of J143753.84+345924.8. The vertical bars show $2\sigma$ 
%%errors. Predicted light curves are overlayed for [Fe/H]~=~-2.0 (solid line); 
%%[Fe/H]~= -1.25 (dot-dash); [Fe/H]~=~-1.0 (long dash); [Fe/H]~=~0.0 (short dash). All four curves are nearly coincident in the ROTSE-I waveband. \label{fig1}}
%%\end{figure}

%%\clearpage

\section{Discussion}

In Figure 1 we compare ROTSE-I and {\it GALEX}  NUV and FUV observed light curves, 
phased using the \citet{ake00} period of 0.564323~$\pm$~0.000058~d.
In the NUV, a variation of AB~=~1.8~mag
is observed, whereas in the FUV this variation is much more pronounced
and reaches $\sim$~5~mag. Measurement errors
for these magnitudes are given in Table 1 and are typically
$\sigma \leq$~0.01~mag at NUV wavelengths and rise to $\sigma = $~0.3~mag
at AB~=~23~mag in the FUV.

In order to model these observed light curves, we require values of
the stellar temperature and radius as a function of phase, together 
with values for stellar metallicity and surface gravity ($\it g$).
Here we assume a mean log~$\it g$ = 2.80, which \citet{ski93}  
has derived for WY Antliae.
Instrument response functions of both the {\it GALEX} \citep{mor04}
and ROTSE-I \citep{woz04,apo04} instruments are required to convert the calculated light 
curves to observed magnitudes. 

The values of stellar temperature
and radius are normally derived using visible and infrared photometry and spectroscopy
using the Baade-Wesselink method. Such empirical data are unfortunately 
not available for R-J14+34. Instead, we use temperature and radius curves derived for  
the well-observed RR Lyrae star WY Antliae \citep{ski93}, which has a period of
 0.574330~d, very close to that of R-J14+34. Furthermore, the 
Two Micron All Sky Survey (2MASS) \citep{cut03}
observed the ROTSE star
near maximum light, when the V-K color of 0.62 was very close to
 the V-K of 0.59 measured by 
\citet{ski93} for WY Antliae at maximum light, a confirmation that the two stars are
physically similar.

We used the model atmosphere grid of \citet{kur03} to
compute the theoretical UV and visible light curves shown in Figure 1. 
These models use the 
updated OPAL opacities of \citet{igl96}, and include the effects
of quasi-molecular absorption by hydrogen \citep{cas01}
that occur near the the center of the {\it GALEX} FUV band. The models,
designated ODFNEW on the Kurucz web-site, incorporate
the ``non-overshoot'' convective treatment of \citet{cas97}. 

In Figure 2 (top panel) we show the Kurucz predicted
UV and visible spectra at minimum and maximum brightness
(i.e. $T_{min}$~=~5900~K and $T_{max}$~=~7300~K). We use Skillen's (1993) 
WY Antliae metallicity of [Fe/H]~=~-1.25 for the model atmosphere. 
In Figure 2 (bottom panel) we show the predicted difference
between maximum and minimum brightness for the 1000 - 9000~\AA\
region.
We note a dramatic increase in amplitude in the
UV region shortward of 1700~\AA\, that rises to AB $>$ 7~mag.
The steep increase in flux between 1500 - 1800~\AA\
is due to the photoionization edge of silicon at 1656~\AA\, and
means that {\it GALEX} FUV observations of RR Lyrae stars are dominated
by 1750~\AA\ photons where the amplitude is AB$\sim$5~mag.
The {\it GALEX} NUV band is dominated by photons at wavelengths
longer than 2700~\AA.

%%\clearpage
%%
%%\begin{figure}
%%\epsscale{0.8}
%%\plotone{f2.eps}
%%\caption{{\it Top panel:} Kurucz model spectra at maximum light (7300~K) and minimum light (5900~K), for [Fe/H] = -1.25 and log~$\it g$  = 2.80. The vertical axis is logarithmic. {\it Bottom panel:} Predicted amplitude of RR Lyrae light curve as a function of wavelength, for [Fe/H] = -2.0 (thick line); [Fe/H] = -1.0 (medium line); [Fe/H] = 0.0 (thin line).\label{fig2}}
%%\end{figure}

%%\clearpage

In Figure 2 (bottom panel) we also show the effect of metallicity
on the amplitude of the predicted light curve. It is immediately
noticeable that the {\it GALEX} NUV and FUV bands are a far more sensitive
indicator of metallicity than either the ROTSE-I visible or
$\lambda <$~1300~\AA\ observations. This effect can also be
seen in Figure 1 in which we have compared the observed
ROTSE-I and {\it GALEX} light curve magnitudes with those derived
from Kurucz model atmospheres using three values of metallicity.
We see that the best fit to the far UV light curve occurs for [Fe/H]~=~-1.25,
which is the value chosen by \citet{ski93} for
the star WY Antliae. The small residual differences between
the observed and predicted UV light curves are probably due to
temperature differences between R-J14+34.5 and WY Antliae.

We emphasize that the derived FUV flux is highly sensitive to a
change in the Kurucz model atmosphere temperature of only
100~-~200~K, particularly around 6000~K at minimum brightness.
A smaller temperature range experienced during the star's oscillation
cycle would require a larger value of metallicity to fit the observed
UV light curves. Our assumption of a constant surface gravity introduces far
smaller errors: varying log~$\it g$ from 2.5 to 3.0 
makes a difference of less than
0.1 mag in the ROTSE-I and GALEX NUV light curves. The largest effect is in the
FUV, where the difference is less than 0.25 mag, compared with the amplitude 
of $\approx$~5 mag, which is largely caused by the temperature variation.    
Other potential sources of error in this analysis
are the uncertainty in the {\it GALEX} FUV magnitudes at AB $>$~22.0~mag
and the uncertainty in the physical parameters derived by \citet{ski93}.

Finally, we note that the high sensitivity and wide field of view
of the {\it GALEX} instrument will result in many more detections
of variable sources during its UV imaging survey of
the sky. In fact, preliminary detections have already been
made for the RR Lyrae stars UU~Indi and HL~Herculis, in which
respective variations in the NUV magnitude of AB~=~2.5~mag and
AB~=~1.8~mag have been recorded.
Kurucz model spectra suggest that 
Cepheids and other instability strip (T~=~6000~-~8000~K) variable
stars should also have large magnitude variations in the FUV.
However, the advantage of observing this large flux variation
in the FUV is unfortunately offset by the corresponding faintness
in the FUV magnitude of these stars. Hence, their potential use as cosmic
distance scale indicators may well be of limited practical value.

\acknowledgments
{\it GALEX} is a NASA Small Explorer, launched in April 2003.
We gratefully acknowledge NASA's support for construction, operation,
and science analysis for the GALEX mission,
developed in cooperation with the Centre National d'Etudes Spatiales
of France and the Korean Ministry of
Science and Technology. The grating, freedom window, and aspheric corrector were supplied by France.
We acknowledge the dedicated
team of engineers, technicians, and administrative staff from JPL/Caltech,
Orbital Sciences Corporation, University
of California, Berkeley, Laboratoire d'Astrophysique de Marseille,
and the other institutions who made this mission possible.
Financial support for this research was provided by NASA grant
NAS5-98034. This publication makes use of data products from the Two Micron All Sky Survey, which is a joint project of the University of Massachusetts and the Infrared Processing and Analysis Center/California Institute of Technology, funded by the National Aeronautics and Space Administration and the National Science Foundation.
We are particularly grateful for the clarifications and improvements suggested by an anonymous referee. 

\begin{figure}
\epsscale{0.8}
\plotone{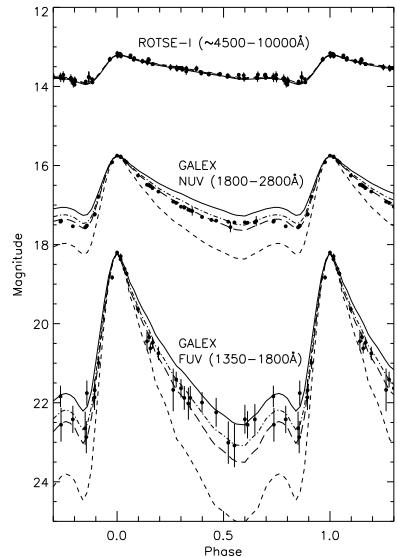}
\caption{Observed ROTSE-I and {\it GALEX} light curves of J143753.84+345924.8 (filled circles). The vertical bars show $2\sigma$ 
errors. Predicted light curves are overlayed for [Fe/H]~=~-2.0 (solid line); 
[Fe/H]~= -1.25 (dot-dash); [Fe/H]~=~-1.0 (long dash); [Fe/H]~=~0.0 (short dash). All four curves are nearly coincident in the ROTSE-I waveband. \label{fig1}}
\end{figure}

\begin{figure}
\epsscale{0.8}
\plotone{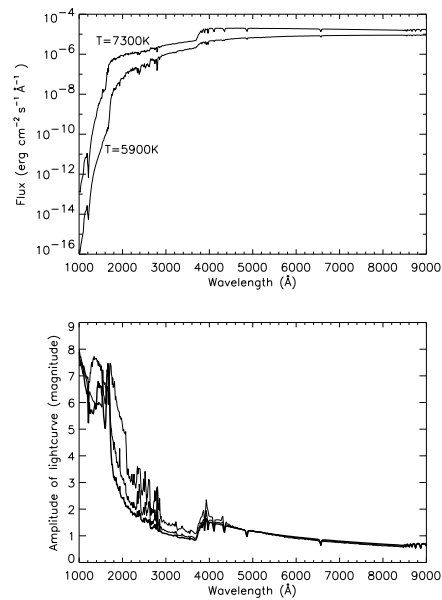}
\caption{{\it Top panel:} Kurucz model spectra at maximum light (7300~K) and minimum light (5900~K), for [Fe/H] = -1.25 and log~$\it g$  = 2.80. The vertical axis is logarithmic. {\it Bottom panel:} Predicted amplitude of RR Lyrae light curve as a function of wavelength, for [Fe/H] = -2.0 (thick line); [Fe/H] = -1.0 (medium line); [Fe/H] = 0.0 (thin line).\label{fig2}}
\end{figure}

\begin{table}
\begin{center}
\caption{{\it GALEX} Photometry of ROTSE-I~J143753.84+345924.8\label{tbl-2}}
\begin{tabular}{rrrrrrrr}
\tableline\tableline
{\it GALEX} Field & Julian Day & Phase\tablenotemark{a} & Exposure & $AB_{NUV}$  & $\sigma_{NUV}$  &  $AB_{FUV}$\tablenotemark{b} & $\sigma_{NUV}$  \\
NGPDWS\_00 & $+$2452800  &  & (s) \\
\tableline

 1  &   5.605636  &  0.443042  &     222  &    17.375  &  0.021  &      -     &   -    \\
 2  &   5.655968  &  0.532231  &      27  &    17.559  &  0.065  &      -     &   -    \\
 3  &   5.724452  &  0.653587  &      25  &    17.405  &  0.063  &      -     &   -    \\
 4  &   5.802598  &  0.792066  &    1693  &    17.537  &  0.008  &    22.421  &  0.172 \\
 5  &   5.871100  &  0.913452  &    1692  &    16.789  &  0.006  &    20.992  &  0.076 \\
 6  &   5.939589  &  0.034818  &    1691  &    15.874  &  0.004  &    18.613  &  0.024 \\
 7  &   6.008085  &  0.156194  &    1691  &    16.522  &  0.005  &    20.625  &  0.064 \\
 8  &   6.076574  &  0.277560  &    1690  &    16.947  &  0.006  &    21.413  &  0.099 \\
 9  &   6.145075  &  0.398947  &    1689  &    17.302  &  0.008  &    21.994  &  0.132 \\
10  &   6.214676  &  0.522281  &    1496  &    17.422  &  0.008  &    23.010  &  0.268 \\
11  &   6.285092  &  0.647062  &    1164  &    17.439  &  0.010  &    22.420  &  0.193 \\
13  &   6.424456  &  0.894019  &     752  &    17.241  &  0.011  &    21.862  &  0.161 \\
14  &   6.493773  &  0.016851  &     608  &    15.788  &  0.006  &    18.296  &  0.035 \\
15  &   6.563513  &  0.140432  &     393  &    16.487  &  0.011  &    20.447  &  0.116 \\
16  &   6.632969  &  0.263511  &     225  &    16.915  &  0.017  &    21.675  &  0.270 \\
17  &   6.683429  &  0.352927  &      24  &    17.151  &  0.057  &      -     &   -    \\
18  &   6.830012  &  0.612678  &    1680  &    17.447  &  0.008  &    22.558  &  0.173 \\
19  &   6.898507  &  0.734054  &    1680  &    17.421  &  0.008  &    21.839  &  0.135 \\
20  &   6.966997  &  0.855420  &    1679  &    17.539  &  0.008  &    22.869  &  0.203 \\
21  &   7.035498  &  0.976806  &    1678  &    15.916  &  0.004  &    18.832  &  0.027 \\
22  &   7.103987  &  0.098172  &    1677  &    16.254  &  0.005  &    19.634  &  0.039 \\
26  &  14.159219  &  0.600288  &    1439  &    17.438  &  0.009  &    22.418  &  0.183 \\
27  &  14.573721  &  0.334800  &     827  &    17.120  &  0.010  &    22.022  &  0.203 \\
27  &  14.573721  &  0.334800  &     827  &    17.120  &  0.010  &    22.022  &  0.203 \\
28  &  15.118443  &  0.300067  &    1383  &    17.042  &  0.007  &    21.634  &  0.117 \\
29  &  15.598090  &  0.150018  &    1354  &    16.502  &  0.006  &    20.338  &  0.062 \\
30  &  16.077755  &  0.000000  &    1322  &    15.751  &  0.004  &    18.202  &  0.023 \\
31  &  16.557448  &  0.850033  &    1289  &    17.565  &  0.010  &    22.646  &  0.210 \\
32  &  17.517454  &  0.551197  &    1134  &    17.440  &  0.010  &    23.081  &  0.274 \\
33  &  18.750671  &  0.736501  &    1148  &    17.399  &  0.010  &    22.556  &  0.214 \\
34  &  18.819213  &  0.857960  &    1144  &    17.518  &  0.010  &    21.757  &  0.159 \\
35  &  19.093409  &  0.343844  &    1131  &    17.138  &  0.009  &    21.878  &  0.151 \\
36  &  19.161956  &  0.465312  &    1126  &    17.394  &  0.010  &    22.243  &  0.183 \\
37  &  19.573252  &  0.194143  &    1106  &    16.659  &  0.007  &    20.753  &  0.085 \\
38  &  19.641800  &  0.315612  &    1103  &    17.066  &  0.008  &    21.878  &  0.151 \\
39  &  20.053119  &  0.044484  &    1089  &    15.917  &  0.005  &    18.734  &  0.032 \\
40  &  20.121673  &  0.165963  &    1087  &    16.571  &  0.007  &    20.486  &  0.075 \\
41  &  20.533009  &  0.894866  &    1078  &    17.228  &  0.009  &    21.682  &  0.140 \\
42  &  20.601557  &  0.016335  &    1075  &    15.772  &  0.005  &    18.372  &  0.027 \\

%%UU Indus 1  &   3.742170 & -       &     741  &    18.632  &   -     &      -     &   -    \\ 
%% 2  &   4.296128 & -       &     533  &    16.135  &   -     &    17.185  &   -    \\

%%HL Her 1  &  21.972662 & -       &   1076   &    18.593  &   -     &    22.099  &   -    \\
%% 2  &  22.246892 & -       &   1083   &    19.492  &   -     &      -     &   -    \\
%% 3  &  22.452502 & -       &   1080   &    17.791  &   -     &    19.506  &   -    \\
%% 4  &  22.792659 & -       &    641   &    19.550  &   -     &      -     &   -    \\

\tableline
\end{tabular}

\pagebreak

%% Any table notes must follow the \end{tabular} command.

\tablenotetext{a}{Phase computed with the ROTSE-I period of 0.564323 $\pm$ 0.000058~d, using {\it GALEX} maximum light at JD 2452816.077755.}
\tablenotetext{b}{A dash indicates that the star was not detected in the FUV observation.}

\end{center}
\end{table}


\clearpage 

\begin{thebibliography}{}

\bibitem[Amrose \& McKay(2001)]{amr01} Amrose, S., \& McKay, T. 2001, \apjl, 560, L151

\bibitem[Akerlof et al.(2000)]{ake00} Akerlof, C., Amrose, S., Balsano, R. et al. 2000, \aj, 119, 1901

\bibitem[Apogee, Inc.(2004)]{apo04} Apogee, Inc. 2004, http://www.ccd.com/AP10.PDF

%%  http://umaxp1.physics.lsa.umich.edu/~mckay/rsv1/rsv1_home.htm

\bibitem[Bertin \& Arnouts(1996)]{ber96} Bertin, E., \& Arnouts, S. 1996, A\&AS, 117, 393

\bibitem[Bohm-Vitense, Proffitt, \& Wallerstein(1984)]{boh84} Bohm-Vitense, Proffitt, C., \& Wallerstein, G.
 1984, in Future of UV Astronomy based on Six Years of IUE Research, NASA Conf. Publ. No. 2349, 348

\bibitem[Bonnell et al.(1982)]{bon82} Bonnell, J., Wu, C-C, Bell, R., \& Hutchinson, J. 1982, \pasp, 94, 910

\bibitem[Bonnell \& Bell(1985)]{bon85} Bonnell, J., \& Bell, R. 1985, \pasp, 97, 236 

\bibitem[Castelli, Gratton, \& Kurucz(1997)]{cas97} Castelli, F., Gratton, R.G., \& Kurucz, R.L. 1997, \aap, 
318, 841

\bibitem[Castelli \& Kurucz(2001)]{cas01} Castelli, F., \& Kurucz, R.L. 2001,
 \aap, 372, 260

\bibitem[Cutri, et al.(2003)]{cut03} Cutri, R.M., et al. 2003, Explanatory Supplement to the 2MASS All Sky Data Release. http://www.ipac.caltech.edu/2mass/releases/allsky/doc/explsup.html

\bibitem[Hutchinson, Hill, \& Lillie(1977)]{hut77} Hutchinson, J., Hill, S., \& Lillie, C. 1977, \apj, 211, 207

\bibitem[Iglesias \& Rogers(1996)]{igl96} Iglesias, C.A., \& Rogers, F.J. 1996, \apj, 464, 943

\bibitem[Kurucz(2002)]{kur02} Kurucz, R.L. 2002, in New Quests in Stellar Astrophysics: The Link between 
Stars and Cosmology, ed. M. Chavez, A. Bressan, A. Buzzoni, \& D. Mayya (Dordrecht: Kluwer), 3 

\bibitem[Kurucz(2003)]{kur03} Kurucz, R.L. 2003, Grids of Model Atmospheres. 
http://kurucz.harvard.edu/grids.html

\bibitem[Martin et al.(2004)]{mar04} Martin, D.C., et al. 2004, \apjl, present volume

\bibitem[McNamara(1997)]{mcn97} McNamara, D.H. 1997, \pasp, 109, 857

\bibitem[Mihalas(2003)]{mih03} Mihalas, D. 2003, in ASP Conf. Ser. 288, Stellar Atmosphere Modeling, 
ed. I. Hubeny, D. Mihalas, \& K. Werner.  ASP Conference Proceedings, 288, 471

\bibitem[Morrissey, et al.(2004)]{mor04} Morrissey, P., et al. 2004, \apjl, present volume

\bibitem[Oke(1974)]{oke74} Oke, J.B. 1974, \apjs, 27, 21

\bibitem[Schlegel, Finkbeiner, \& Davis(1998)]{sch98} Schlegel, D., Finkbeiner, D., \& Davis, M., 1998, \apj, 500, 525

\bibitem[Skillen et al.(1993)]{ski93} Skillen, I., Fernley, J., Stobie, R., \& Jameson, R. 1993, \mnras, 
265, 301

\bibitem[Wozniak et al.(2004)]{woz04} Wozniak, P.R., Vestrand, W.T., Akerlof, C.W. et al. 2004, \aj, 127, 2436  

\end{thebibliography}
\end{document}